%% file: main.tex
\documentclass[a4paper,twocolumn,11pt,accepted=2024-02-13]{quantumarticle}
\pdfoutput=1
\usepackage[english]{babel}
\usepackage[utf8]{inputenc}
\usepackage[unicode=true, colorlinks=true]{hyperref} 
\input{preamble}
\begin{document}
\title{Variational Phase Estimation with Variational Fast Forwarding}

\author{Maria-Andreea Filip}
    \affiliation{Quantinuum, 13-15 Hills Road, CB2 1NL,  Cambridge, United Kingdom}
    \affiliation{Yusuf Hamied Department of Chemistry, University of Cambridge, Cambridge, United Kingdom}

\author{David Mu\~{n}oz Ramo}
    \affiliation{Quantinuum, 13-15 Hills Road, CB2 1NL, Cambridge, United Kingdom}

\author{Nathan Fitzpatrick}
    \email{nathan.fitzpatrick@quantinuum.com}% Your name
    \affiliation{Quantinuum, 13-15 Hills Road, CB2 1NL, Cambridge, United Kingdom}

\maketitle
\begin{abstract}
Subspace diagonalisation methods have appeared recently as promising means to access the ground state and some
excited states of molecular Hamiltonians by classically diagonalising small matrices, whose elements can
be efficiently obtained by a quantum computer. The recently proposed Variational Quantum Phase Estimation (VQPE)
algorithm uses a basis of real time-evolved states, for which the energy eigenvalues can be obtained directly
from the unitary matrix $\mathbf{U} = e^{-i\mathbf{H}\Delta t}$, which can be computed with cost linear in the
number of states used. In this paper, we report a circuit-based implementation of VQPE for arbitrary molecular
systems and assess its performance and costs for the H$_2$, H$_3^+$, H$_6$ and LiH molecules. We also propose using
Variational Fast Forwarding (VFF) to decrease to quantum depth of time-evolution circuits for use in VQPE. We show
that the approximation provides a good basis for Hamiltonian diagonalisation even when its fidelity to the true time
evolved states is low. In the high fidelity case, we show that the approximate unitary $\mathbf{U}$ can be diagonalised
instead, preserving the linear cost of exact VQPE.
\end{abstract}

\maketitle

\input{sections/introduction.tex}

\input{sections/methods.tex}
\input{sections/results.tex}

\input{sections/conclusion.tex}

\input{sections/acknowledgements.tex}

\bibliographystyle{quantum}
\bibliography{references2.bib}
%\printbibliography
\input{sections/appendices}
\onecolumngrid

\end{document}

%% file: preamble.tex
\usepackage{amsthm}
\usepackage{amsmath}
\usepackage{amsfonts}
\usepackage{amssymb}
\usepackage{mathtools}
\usepackage{physics}
\usepackage{xcolor}
\usepackage{graphicx}
\usepackage[left=23mm,right=13mm,top=35mm,columnsep=15pt]{geometry} 
\usepackage{adjustbox}
\usepackage{placeins}
\usepackage[T1]{fontenc}
\usepackage{lipsum}
\usepackage{csquotes}
\usepackage{tikz}
\usetikzlibrary{quantikz}
\usepackage{float}
\usepackage{bm}
\usepackage{braket}
\usepackage[capitalise]{cleveref}
\usepackage{multirow}
\usepackage{comment}
\PassOptionsToPackage{compress}{natbib}
\usepackage[numbers]{natbib}

%% file: sections/introduction.tex
\section{Introduction} \label{sec:introduction}

The current era of noisy intermediate scale quantum (NISQ) devices\cite{Preskill2018} is dominated by hybrid quantum-classical algorithms like the variational quantum eigensolver (VQE).\cite{Peruzzo2014} In this approach a cost function is calculated on a quantum computer and then minimised using a classical optimisation procedure, with the aim of reducing the amount of quantum computation to a level manageable on current architectures, built of noisy gates and qubits with low coherence times. VQE has found use in quantum chemistry applications,\cite{O'Malley2016, Hempel2018} where the goal often is finding estimates of the lowest eigenvalue(s) of the molecular Hamiltonian. Alternative hybrid approaches to this problem have also been developed, based on the imaginary-time evolution of the system\cite{McArdle2019} and, more recently, on real-time evolution.\cite{Parrish2019}

Quantum Phase Estimation (QPE)\cite{Kitaev1995} --- which uses the repeated application of a unitary and the quantum Fourier transform to estimate the phase of the unitary --- was one of the first quantum algorithms highlighted for quantum chemistry, by using it on the real-time evolution operator to get the lowest eigenvalue of the Hamiltonian.\cite{Aspuru-Guzik2005} However, its quantum resource requirements make it intractable on NISQ devices. The recently developed Variational Quantum Phase Estimation (VQPE) algorithm\cite{Klymko2022} uses time evolved states as a basis for non-orthogonal configuration interaction (NOCI). The NOCI matrix elements are computed on the quantum processor, while the diagonalisation is carried out classically. It has been shown that only relatively few states and a linear number of measurements are needed.

This algorithm is part of a wider class of quantum subspace diagonalisation (QSD) methods,\cite{McClean2017,Parrish2019,Huggins2020, Motta2020, Stair2020, Cortes2022} in which non-orthogonal states are used as a basis
to solve the generalised Hamiltonian eigenvalue problem, giving estimates for multiple eigenstates. Many of these methods, including VQPE, perform the diagonalisation in a Krylov subspace\cite{Golub1983} obtained by the repeated application of some unitary.\cite{Stair2020,Cortes2022}

In this paper, we describe a quantum circuit based implementation of the VQPE algorithm for general molecular Hamiltonians and assess the hardware requirements of such a method. We also investigate using Variational Fast Forwarding (VFF)\cite{Cirstoiu2020, Gibbs2021} to approximate time-evolution and the effects this approximation has on the resulting VQPE calculation.

We begin with an overview of the existing VQPE algorithm in Sec. \ref{sec:back}. In Sec. \ref{sec:methods}, we describe our novel circuit implementation of VQPE and the subsequent VFF approximation, while Sec. \ref{sec:results} shows the results of applying these methods to the Hubbard model, small linear H$_n$ species and LiH. Conclusions are drawn in Sec. \ref{sec:conclusion}.

\section{Background}\label{sec:back}

\subsection{Variational Quantum Phase Estimation} \label{sec:vqpe}

Like the standard phase estimation algorithm (see \cref{sec:qpe} for a comparison), Variational Phase Estimation (VPE) requires a reference state with some non-zero overlap with the eigenbasis of the Hamiltonian.\cite{Klymko2022} We can expand the reference state $\ket{\Phi_0}$ in terms of the eigenfunctions $\ket{N}$ as
\begin{equation}
    \ket{\Phi_0} = \sum_N^Q \phi^0_N\ket{N},
\end{equation}
where $\phi^0_N = \braket{N | \Phi_0}$. The key step in VQPE is to define a series of time-evolved states $\ket{\Phi_{j,0}}$,
\begin{equation}
    \ket{\Phi_{j,0}} = e^{-i\hat Ht_j}\ket{\Phi_0},
\end{equation}
which can also be expanded in the eigenbasis of $\hat H$ as
\begin{equation}
    \ket{\Phi_{j,0}} = \sum_N^Q\phi^0_N e^{-iE_Nt_j}\ket{N}.
    \label{eq:support}
\end{equation}
These time-evolved states form a Krylov subspace which may be used as a basis for configuration interaction (CI) or exact diagonalisation (ED) \cite{zbMATH02561215}. Consider a wavefunction
\begin{equation}
\ket{\Psi} = \sum_{j = 0}^{N_T} c_j \ket{\Phi_{j,0}}.
\label{eq:linear}
\end{equation}
Minimising $\braket{E}$ with respect to all $c_j$ for this wavefunction according to the variational principle is equivalent to solving the generalised eigenvalue problem
\begin{equation}
    \mathbf{Hc} = \epsilon\mathbf{Sc}.
    \label{eq:gen_eig}
\end{equation}
where the elements of matrices $\mathbf{H}$ and $\mathbf{S}$ are given by
\begin{equation}
    H_{jk} = \braket{\Phi_{j,0}|\hat H|\Phi_{k,0}}
\end{equation}
and
\begin{equation}
    S_{jk} = \braket{\Phi_{j,0}|\Phi_{k,0}} = \braket{\Phi_{0} | e^{-i \hat H ( t_k - t_j}) | \Phi_{0}}
\end{equation}
respectively. Solving this matrix equation gives approximate values for $N_T + 1$ eigenvalues of $\hat H$.

Consider the overlap operator
\begin{align}
\begin{split}
    \hat S &= \sum_{j=0}^{N_T} \ket{\Phi_{j,0}}\bra{\Phi_{j,0}}\\
      &= \sum_{N,M}^Q \phi^0_N \phi^{0*}_M\Bigg[\sum_{j=0}^{N_T} e^{-it_j(E_N-E_M)}\Bigg] \ket{N}\bra{M}.
\end{split}
\end{align}
The original work by Klymko \textit{et al.}\cite{Klymko2022} defines a set of phase cancellation conditions,
\begin{equation}
    \frac{1}{N_T+1}\sum_{j=0}^{N_T} e^{-it_j(E_N-E_M)} = \delta_{N,M}
\end{equation} 
under which 
\begin{equation}
\hat S = (N_T + 1) \sum_N^Q |\phi^0_N|^2 \ket{N}\bra{N}.
\end{equation}
If these conditions are satisfied then the set of time-evolved states will span
the full support space of the initial state and VQPE will be able to perfectly recover all the eigenstates in that support space. This requires at least as many time-evolved states as there are eigenstates in the support space, which may not be tractable for general systems. However, even an incomplete basis may give a good approximation for the ground state energy of the system. { Furthermore, the fact that time-evolution does not introduce any additional eigenstates beyond those present in the initial state (as can be seen from \cref{eq:support}) guarantees that a non-increasing number of basis states is needed to resolve those eigenvectors.} For traditional configuration interaction algorithms like configuration interaction with single and double excitations (CISD), there are no such guarantees in place.

{It has also been shown\cite{Klymko2022} that an alternative} to \cref{eq:gen_eig} may be defined by considering the time evolution operator $\hat U(\Delta t) = e^{-i\hat H\Delta t}$. For eigenfunctions of the Hamiltonian,
\begin{equation}
    \hat U(\Delta t) \ket{N} = e^{-iE_N\Delta t}\ket{N}.
\end{equation}
Therefore we can define an alternate generalised eigenvalue problem, 
\begin{equation}
    \mathbf{U}(\Delta t)\mathbf{c} = \lambda\mathbf{Sc}.
    \label{eq:gen_eig2}
\end{equation}

\begin{figure*}
\centering
\begin{quantikz}
\lstick{$|a\rangle$} & \qw & \qw & \qw & \qw &\ctrl{4} & \qw & \qw & \qw&\qw & \qw \\
\lstick{$|i_3\rangle$} & \gate{R_x(\frac{3\pi}{2})} & \ctrl{1} & \qw & \qw &\qw & \qw & \qw & \ctrl{1} &\gate{R_x(\frac{\pi}{2})} & \qw \\
\lstick{$|i_2\rangle$} & \gate{H} &\targ{}  & \ctrl{1} & \qw & \qw & \qw & \ctrl{1}  & \targ{} & \gate{H} & \qw \\
\lstick{$|i_1\rangle$} & \gate{H} &\qw & \targ{}  & \ctrl{1} & \qw & \ctrl{1} & \targ{} & \qw & \gate{H} & \qw \\
\lstick{$|i_0\rangle$} & \qw &\qw & \qw & \targ{}  & \gate{R_z(\theta)} & \targ{}  & \qw & \qw & \qw & \qw \\
\end{quantikz}
\caption{\small Controlled Pauli gadget for $e^{-i\frac{\theta}{2}(Z_0 \otimes X_1 \otimes X_2 \otimes Y_3)}$.}
\label{fig:c-gadget}
\end{figure*}
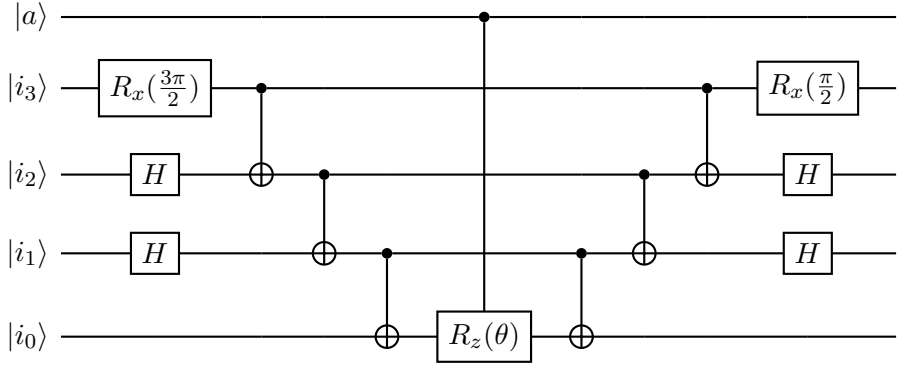

The solutions to \cref{eq:gen_eig,eq:gen_eig2} are related in the same way as the corresponding operators, up to a phase factor:
\begin{equation}
    \epsilon_N = -\frac{\mathrm{log}(\lambda_N)}{i\Delta t} \pm \frac{2n\pi}{\Delta t}.
\end{equation}
The time evolution matrix elements are given by

\begin{align}
\begin{split}
U_{jk}  &= \braket{\Phi_{j,0}| e^{-i \hat H \Delta t}|\Phi_{k,0}}\\
&= \braket{\Phi_{0} | e^{-i \hat H (\Delta t + t_k - t_j}) | \Phi_{0}}
\end{split}
\label{eq:U_def}
\end{align}

In the case where a uniform time grid $t_j = j \Delta t$ is used, these coincide with elements of the overlap matrix
\begin{equation}
    U_{j,k} = S_{j,k+1} = S_{j-1,k}.
    \label{eq:U_S_relation}
\end{equation}
Therefore, while \cref{eq:gen_eig} requires separate measurements of all Hamiltonian and overlap matrix elements, for \cref{eq:gen_eig2} one only needs to measure the elements of the overlap matrix to obtain both $\mathbf{U}$ and $\mathbf{S}$. Furthermore, in the uniform time grid case
\begin{align}
\begin{split}
    S_{jk} &=  \braket{\Phi_{0} | e^{-i \hat H ( t_k - t_j)} | \Phi_{0}}\\
           &=  \braket{\Phi_{0} | e^{-i \hat H \Delta t(k - j)} | \Phi_{0}}\\
           &= S_{j+n,k+n},\ \forall n \in \mathbb{Z}
    \end{split}
    \label{eq:S_struct}
\end{align}
Therefore, to construct the full overlap matrix, one only needs to compute one row of elements. This is a significant advantage to this algorithm, as it reduces the number of measurements to be linear in the number of time-evolved states rather than quadratic. As in direct Hamiltonian diagonalisation, the eigenvalues of $\hat U$ provide estimates for both the ground and excited states of the system, although the presence of phase factors may confuse the identity of the states. Finding the eigenvectors of $\hat U$ would then give the variationally optimised parameters $c_j$ in \cref{eq:linear}.

%% file: sections/methods.tex
\section{Methods}\label{sec:methods}
\subsection{Circuit Implementation of VQPE} \label{sec:circuit}

\begin{figure}
\centering
\vspace{10pt}
\begin{quantikz}
\lstick{$\ket{a}$} &  \gate{H}& \ctrl{1} &\gate{X} &\ctrl{1}&\gate{X}&\gate{H}&\qw\\
\lstick{$\ket{\Phi_0}$}&\qwbundle[alternate]{} &\gate{U^j}\qwbundle[alternate]{}&\qwbundle[alternate]{}&\gate{U^k} \qwbundle[alternate]{}&\qwbundle[alternate]{}&\qwbundle[alternate]{}&\qwbundle[alternate]{}
\end{quantikz}
\caption{Measurement circuit for $S_{jk} = \braket{\Phi_{j,0}|\Phi_{k,0}} = \braket{\Phi_0|\hat (U^\dagger)^j \hat U^k |\Phi_0}$.}
\label{fig:overlap}
\end{figure}
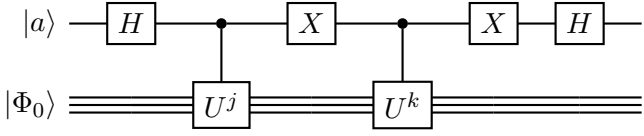

In second quantisation, a chemistry Hamiltonian may be written as
\begin{equation}
    \hat H = \sum_{p,q} f_{pq}\hat a_p^\dagger \hat a_q + \sum_{p,q,r,s} h_{pqrs} \hat a_p^\dagger \hat a_q^\dagger \hat a_r \hat a_s,
\end{equation}
where $\hat a_p^\dagger$ and $\hat a_p$ are fermionic creation and annihilation operators respectively. These operators may be mapped onto strings of Pauli operators acting on qubits by a variety of schemes such as the Jordan-Wigner\cite{Jordan1928} or Bravyi-Kitaev\cite{Bravyi2002} encodings. For example, in the Jordan-Wigner mapping
\begin{equation}
\hat a^\dagger_j = \bigotimes^{j-1}_i Z_i \bigotimes \frac{1}{2}(X_j - iY_j),
\end{equation}

\begin{equation}
\hat a_j = \bigotimes^{j-1}_i Z_i \bigotimes \frac{1}{2}(X_j + iY_j).
\end{equation}
By applying this transformation to all terms in the Hamiltonian, we obtain the equivalent qubit Hamiltonian
\begin{equation}
    \hat H = \sum_k \hat h_k  = \sum_k h_k \hat P_k
\end{equation}
where $\hat P_k$ are strings of Pauli operators. One can then take the first order Trotter-Suzuki approximation to the time evolution operator, for a finite time step of the whole evolution. 
\begin{equation}
e^{-i\hat H\Delta t} \approx \prod_{k} e^{-i\hat h_k\Delta t} = \hat U.
\label{eq:trot}
\end{equation}
This is exact in the limit of an infinitely small time step. Trotterization can be efficiently encoded in a quantum circuit as a Pauli gadget\cite{Cowtan2020} (see \Cref{fig:c-gadget}) and propagation by multiple time-steps may then be approximated by
\begin{equation}
e^{-i\hat Hj\Delta t} \approx \hat U^j.
\end{equation}
We therefore encode $\ket{\Phi_{j,0}}$ as
\begin{equation}
    \ket{\Phi_{j,0}} = \hat U^j \ket{\Phi_0}
\end{equation}
where we chose $\ket{\Phi_0}$ to be the Hartree--Fock wavefunction for the system. In order to compute $S_{jk}$ we implement the circuit in \cref{fig:overlap}.

At the end of the circuit, the qubit register is in the state
\begin{equation}
\ket{\Psi} = \frac{1}{2}\ket{0}(\ket{\Phi_{k,0}} + \ket{\Phi_{j,0}}) + \frac{1}{2}\ket{1}(\ket{\Phi_{k,0}} -\ket{\Phi_{j,0}})
\end{equation}
and measuring the ancilla in either the $Z$ or $Y$ basis gives the real and imaginary parts of $S_{jk}$ respectively.

\begin{align}
\begin{split}
\braket{\Psi|Z_a|\Psi} &= \mathrm{Re}(\braket{\Phi_{j,0}|\Phi_{k,0}}) \\
\braket{\Psi|Y_a|\Psi} &= \mathrm{Im}(\braket{\Phi_{j,0}|\Phi_{k,0}})
\end{split}
\end{align}
Additionally,
\begin{align}
\begin{split}
\braket{\Psi|Z_a \hat H|\Psi} &= \mathrm{Re}(\braket{\Phi_{j,0}|\hat H|\Phi_{k,0}}) \\
\braket{\Psi|Y_a \hat H|\Psi} &= \mathrm{Im}(\braket{\Phi_{j,0}|\hat H|\Phi_{k,0}})
\end{split}
\end{align}
so Hamiltonian elements may be computed using the same circuit if needed.

{In the case of Pauli gadgets, it is easy to implement the controlled version, which is also used for imaginary time propagation.\cite{chan2023simulating} Because the gadget encodes a similarity transformation of the form $\hat G \hat R_z(\theta) \hat G^\dagger$,  the controlled operation only requires controlling the central $R_z$ gate as can be seen in figure \cref{fig:c-gadget}.} For number conserving $\hat U$ operators, alternative measurement circuits are possible which do not require controlled operations.\cite{Huggins2020} We have found that because individual Pauli strings in the trotterised time-evolution operator are not number preserving, these measurement techniques do not perform well in this case. If \cref{eq:S_struct} holds, we only need to compute the first row elements of the overlap, $S_{0k}$, so the circuit in \cref{fig:overlap} simplifies to a Hadamard test.

\subsection{Approximating VQPE with Variational Fast Forwarding}\label{sec:vff}

\begin{figure*}
    \centering
    \includegraphics[width=\textwidth]{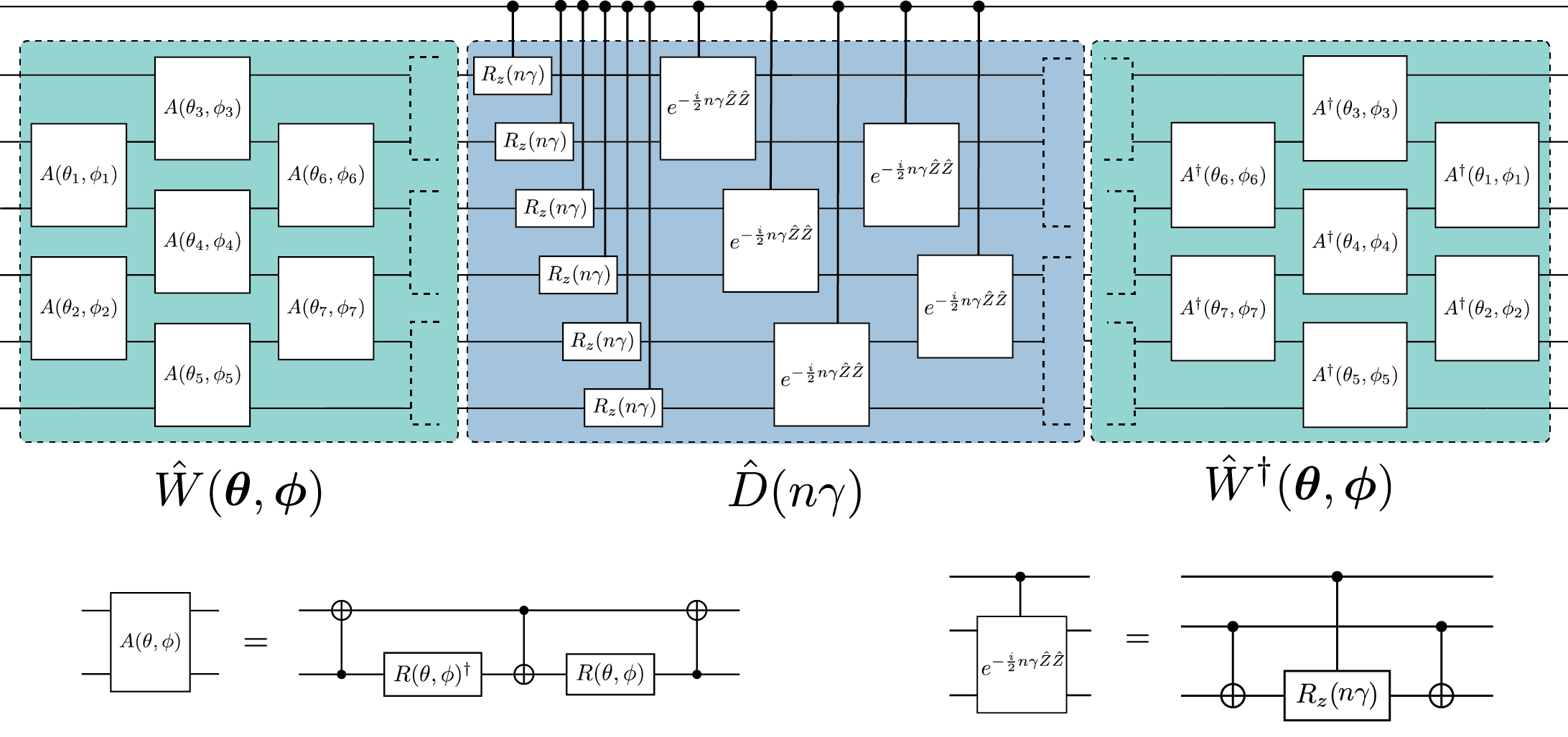}
    \caption{Controlled time evolution operator compiled using variational fast forwarding and number conserving gates.  This primitive is used to calculate the overlap $\braket{\Phi_0|\hat W \hat D^{k-j}\hat W^\dagger|\Phi_0} $ using the Hadamard test protocol.}
    \label{fig:cnt_VFF}
\end{figure*}

As we show in Sec. \ref{sec:results}, the depth of Trotterised time-evolution circuits increases intractably, so we investigate potential routes to circumvent this problem. One approach, which leads to constant-depth approximations for time-evolution circuits, is Variational Fast Forwarding (VFF), \cite{Cirstoiu2020, Gibbs2021} in which the time-evolution operator is approximated as
\begin{equation}
    e^{-i\hat H \Delta t} \approx \hat V(\boldsymbol{\theta}, \boldsymbol{\gamma}) \approx \hat W(\boldsymbol{\theta}) \hat D(\boldsymbol{\gamma}) \hat W^\dagger(\boldsymbol{\theta}),
    \label{eq:vff}
\end{equation}
where $\hat D$ is a diagonal operator and $\hat W$ is an arbitrary unitary operator. In this case
\begin{equation}
    e^{-i\hat H n\Delta t} \approx \hat W \hat D^n \hat W^\dagger.
\end{equation}
The diagonal part may be expressed as
\begin{equation}
    \hat D(\boldsymbol{\gamma}; \Delta t) = \prod_{m=0}^{n-1} \prod_{j \in S_m} e^{i\gamma_j \bigotimes_{k=0}^{n-1}(Z_k)^{j_k}},
\end{equation}
where $S_m$ is the set of $n$-bit binary numbers with $m$ bits set, $j_k$ is the value of the $k^\mathrm{th}$ bit in $j$ and $\boldsymbol{\gamma}$ is a set of parameters to be variationally optimised. This expression may be further approximated by truncating $m$. In this parametrisation, 
\begin{equation}
    \hat D^n(\boldsymbol{\gamma}; \Delta t) = \hat D(n\boldsymbol{\gamma}; \Delta t),
\end{equation}
making the construction of operators for different values of $\Delta t$ easy and cheap. The $\hat W$ operator can be parametrised in various ways, provided they have enough flexibility to express the required unitary. Different layered ansatze have been used for this\cite{Cirstoiu2020,Gibbs2021} and in this work we use the symmetry preserving {ansatz},\cite{Gard2020} which is number preserving and can also be designed to avoid symmetry and spin contamination. The circuit structure for the controlled evolution operator can be seen in figure \cref{fig:cnt_VFF}.

In order to optimise the parameters $\boldsymbol{\theta}$ and $\boldsymbol{\gamma}$ a cost function of the form\cite{Gibbs2021} 
\begin{equation}
f(\boldsymbol{\theta},\boldsymbol{\gamma}) = 1 - \frac{1}{n}\sum_{k=1}^n |\braket{\psi|(V^k)^\dagger e^{-iHk\Delta t}|\psi}|^2
\label{eq:cost_func}
\end{equation}
is minimised. The cost function is 0 when the overlap between the states generated by time-evolution and those generated by repeatedly applying $\hat V$ is perfect. Exact time-evolution will preserve the support space of the initial wavefunction $\ket{\psi}$, so the unitary $\hat V$ must match the action
of the time-evolution operator in this space. The No-Free-Lunch theorem \cite{Poland2020} then dictates that, in order to fully describe the action of the time-evolution unitary on the support space of the initial state, the cost function would require one term for each dimension in this space.\cite{Gibbs2021} 

This approach is hardly useful if we are trying to devise a lower-cost alternative to VQPE, as it would require measurements of as many, if not more, overlaps with time-evolved states as the full VQPE approach to obtain and optimise the cost-function, thereby not removing the need for extremely long Trotter circuits.

Instead, we relax the condition that the non-orthogonal basis must be constructed from exact time-evolved states. As long as $\hat U$ commutes with the Hamiltonian it will share the same eigenfunctions, giving
\begin{equation}
   \hat U \ket{\Phi_0} = \sum_N \phi^0_N\hat U\ket{N} = \sum_N \phi^0_N u_N\ket{N},
\end{equation}
where $u_N$ are the corresponding eigenvalues of $\hat U$.

The repeated application of $\hat U$ does not change the support space of $\ket{\Phi_0}$, so it could be used instead of the time-evolution operator. Unitary operators that do not commute with the Hamiltonian can also be used to generate basis states for exact diagonalisation, however they are not guaranteed to maintain the same support space for all states and therefore may require more basis functions to obtain good estimates of the true eigenstates. 

For $\hat U \neq e^{-i\hat H \Delta t}$, there is no longer an explicit relationship between the eigenvalues of the Hamiltonian and those of $\hat U$, so the former cannot be easily extracted from the diagonalisation of $\mathbf{U}$. If $[\hat U, \hat H] = 0$, then the Hamiltonian may be applied to the eigenfunctions of $\hat U$ to obtain the eigenvalues of $\hat H$. If $[\hat U, \hat H] \neq 0$ however, the only option is to construct the Hamiltonian matrix in the basis of states obtained by repeated application of $\hat U$ and directly diagonalise it. 

We also note that, regardless of the relationship of $\hat U$ with the time-evolution operator, this algorithm amounts to a Krylov subspace diagonalisation, where the space is given by
$\{\ket{\Phi_0}, \hat U \ket{\Phi_0}, \hat U^2 \ket{\Phi_0}...\}$. Provided enough linearly independent states are considered to span the full support space of this Krylov subspace, the eigenvalues obtained from the diagonalisation should be correct. Therefore we expect that even a poor VFF approximation to $e^{-i\hat H \Delta t}$ could be employed in VQPE and still generate good results, if direct Hamiltonian diagonalisation is used. 

{ One of the significant advantages of VQPE is that, in the basis of time-evolved states, the overlap matrix $\mathbf{S}$ can be computed with linear cost due to \cref{eq:S_struct}. This remains rigorously the case for VFF approximate states, for which
\begin{align}
    \begin{split}
    S_{jk} &= \braket{\Phi_0|\hat W \hat D^{-j} \hat W^\dagger \hat W \hat D^{k} \hat W^\dagger|\Phi_0} \\
    &= \braket{\Phi_0|\hat W \hat D^{-j}  \hat D^{k} \hat W^\dagger|\Phi_0}\\
    &= \braket{\Phi_0|\hat W \hat D^{k-j}\hat W^\dagger|\Phi_0} \\
    &= S_{j+n,k+n}, \forall n \in \mathbb{Z}
    \end{split}
\end{align}
This is particularly useful when applying \cref{eq:gen_eig2}, as it reduces the cost of the computation of $\mathbf{U}$ to linear as well. In the case of VFF, while Hamiltonian diagonalisation is resilient to a poor VFF approximation as its quality depends only on a basis set's ability to span the relevant support space, that is not the case for \cref{eq:gen_eig2} with $\mathbf{U}$ defined as in \cref{eq:U_S_relation}. If the VFF states are poor approximations to the true time-evolved states, 
\begin{equation}
    U_{jk} = S_{j, k+1} \neq \braket{\Phi_j|e^{-i\hat H \Delta t}|\Phi_k}
\end{equation}
so the eigenvalues of $\mathbf{U}$ are no longer clearly related to the Hamiltonian eigenvalues and, in particular, the method is no longer guaranteed to be variational. However, if the approximation is good enough, it should be possible to employ this approach.}

%% file: sections/results.tex
\section{Results}\label{sec:results}

We showcase the VPE and VFF-VPE methods on a range of systems including the two-site Hubbard model, hydrogen systems (H$_2$, { H$_3^+$ in a linear geometry and the H$_6$ chain}) and LiH using the Qiskit Aer shot-based quantum simulator\cite{Qiskit} and comparing to results obtained by direct matrix algebra on the corresponding state-vectors. { We apply (VFF-)VPE to these systems to obtain binding curves describing the behaviour of the total ground state energy $E$ (reported here in units of Hartree $E_h$) as the distances between atoms are increased. We also investigate the convergence behaviour of the calculations with time-grid size $\Delta t$, number of time-evolved states $N_T$ and quality of the VFF approximation.} 

\subsection{Implementation}
All following implementation and calculations were carried out using the InQuanto\cite{InQuanto} framework. 

\textbf{VFF Parameter Optimisation.} We initialise the system in the Hartree--Fock state $\ket{\Phi_0}$ and  generate the first few exact time-evolved states using a state-vector simulation of the Hamiltonian. We also obtain the parameterised state-vector representation of the VFF wavefunctions corresponding to the same times by applying the circuit in \cref{fig:cnt_VFF} to the same inital state. We classically compute their overlap and the resulting value of the cost function according to \cref{eq:cost_func}. This is then optimised using the L-BFGS algorithm\cite{Liu1989} to generate the final VFF unitary. In principle, it may be possible to do this type of variational compilation on a quantum device, provided the Hamiltonian is local.\cite{Mizuta2022}  

\textbf{Time-evolution and diagonalisation.} For both trotterised time-evolution and VFF, states are then constructed by repeated application of the quantum circuit for unitary $\hat U$ from \cref{eq:trot} or \cref{eq:vff} respectively. The overlaps and Hamiltonian elements between time-evolved states is obtained by using the circuit in \cref{fig:overlap}. The Hamiltonian or $\mathbf{U}$ matrix is projected into the space of linearly independent eigenvectors of $\mathbf{S}$ and diagonalised. The resulting eigenstates can be projected back into the full space of time-evolved states to obtain a representation of the corresponding wavefuntion in this basis. 

\textbf{Shot-based and state-vector simulation.} For shot-based simulation all measured quantities were obtained using 10000 shots. A linear dependecy threshold of $0.1$ on the eigenvalues of the overlap matrix was used to define the number of linearly independent vectors in the space. { This agrees with the value chosen by Klymko \textit{et al}\cite{Klymko2022} and is sufficiently larger than the expected finite sampling noise to avoid spurious states.} While lower thresholds could be used for some systems, leading to faster convergence, a large value is more consistent with the high noise we expect from real hardware. Multiple independent simulations are averaged over to obtain the final results and error bars.
For state-vector simulation, a much lower linear dependency threshold ($10^{-5}$) was used.

\subsection{Binding curves}

\begin{figure}
    \centering
    \includegraphics[width=0.5\textwidth, trim = 0 0 0 1cm, clip]{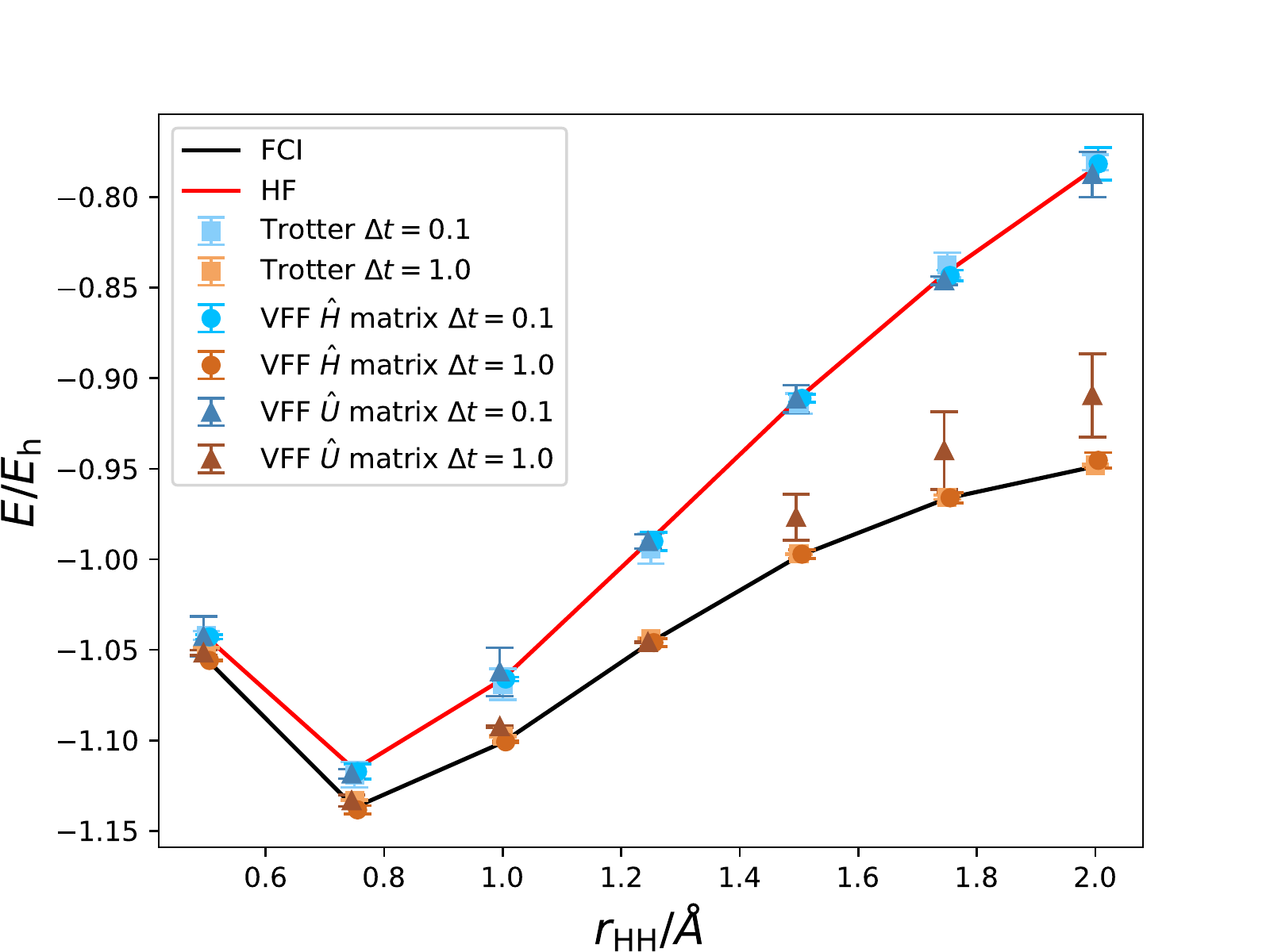}
    \includegraphics[width=0.5\textwidth, trim = 0 0 0 1cm, clip]{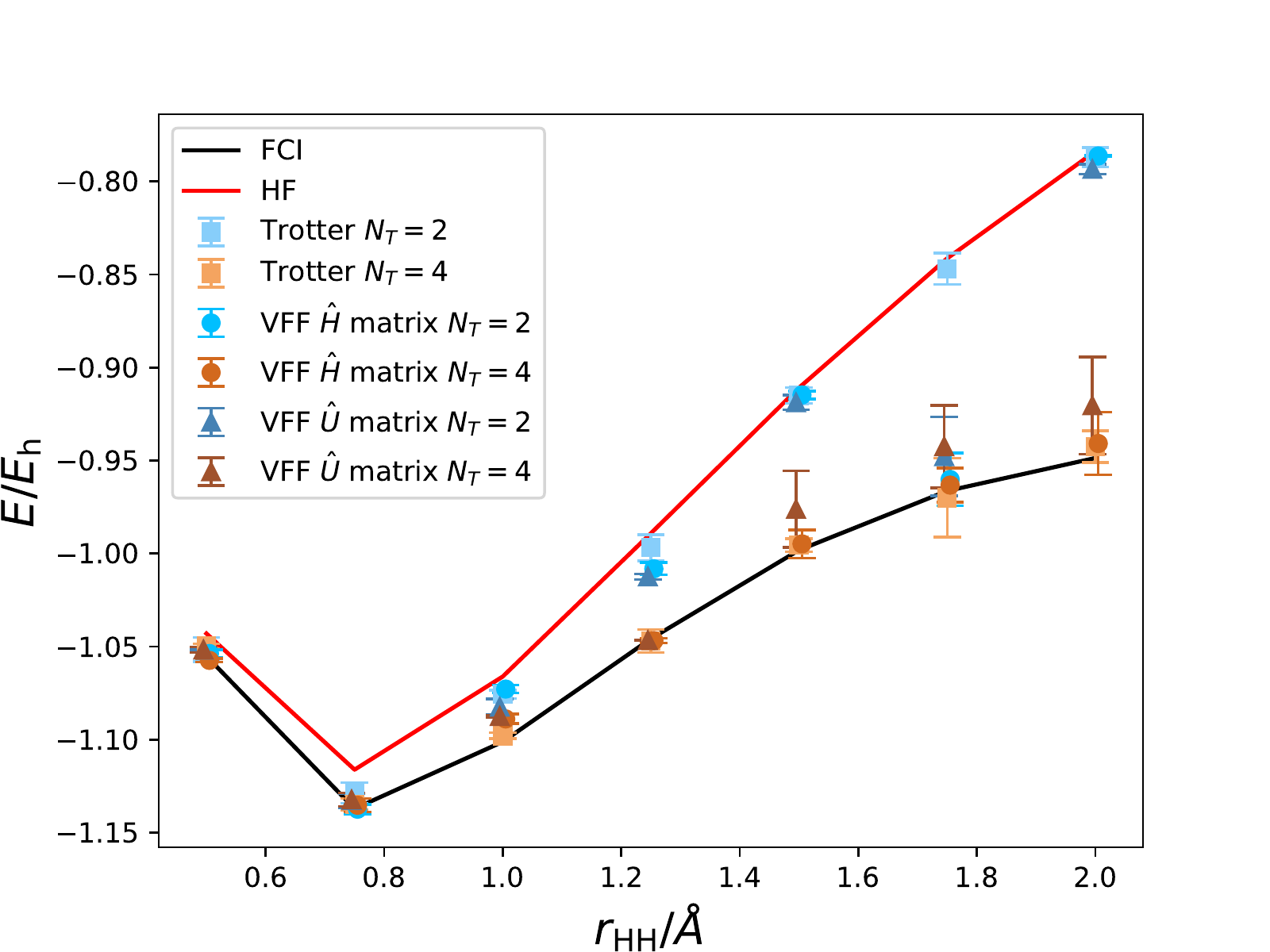}
    \caption{ \small H$_2$ binding curve in the STO-3G basis obtained from shot-based VQPE simulations, with or without the VFF approximation, using $N_T = 10$ and varying sizes of time grid (top) and $\Delta t = 1.0$ and varying numbers of basis states (bottom). Error bars were obtained by averaging results over multiple different simulations. VFF $\hat H$ matrix results are obtained using \cref{eq:gen_eig} and $\hat U$ matrix results with \cref{eq:gen_eig2}. Sufficiently large $\Delta t$ and $N_T$ are required to introduce new linearly independent states and therefore obtain a lower-than-HF energy from VQPE. As H$_2$ has a two-dimensional Hilbert space, only one additional state must be obtained match FCI.}
    \label{fig:H2}
\end{figure}

\begin{figure}[h!]
    \centering
    \includegraphics[width=0.5\textwidth, trim = 0 0 0 1cm, clip]{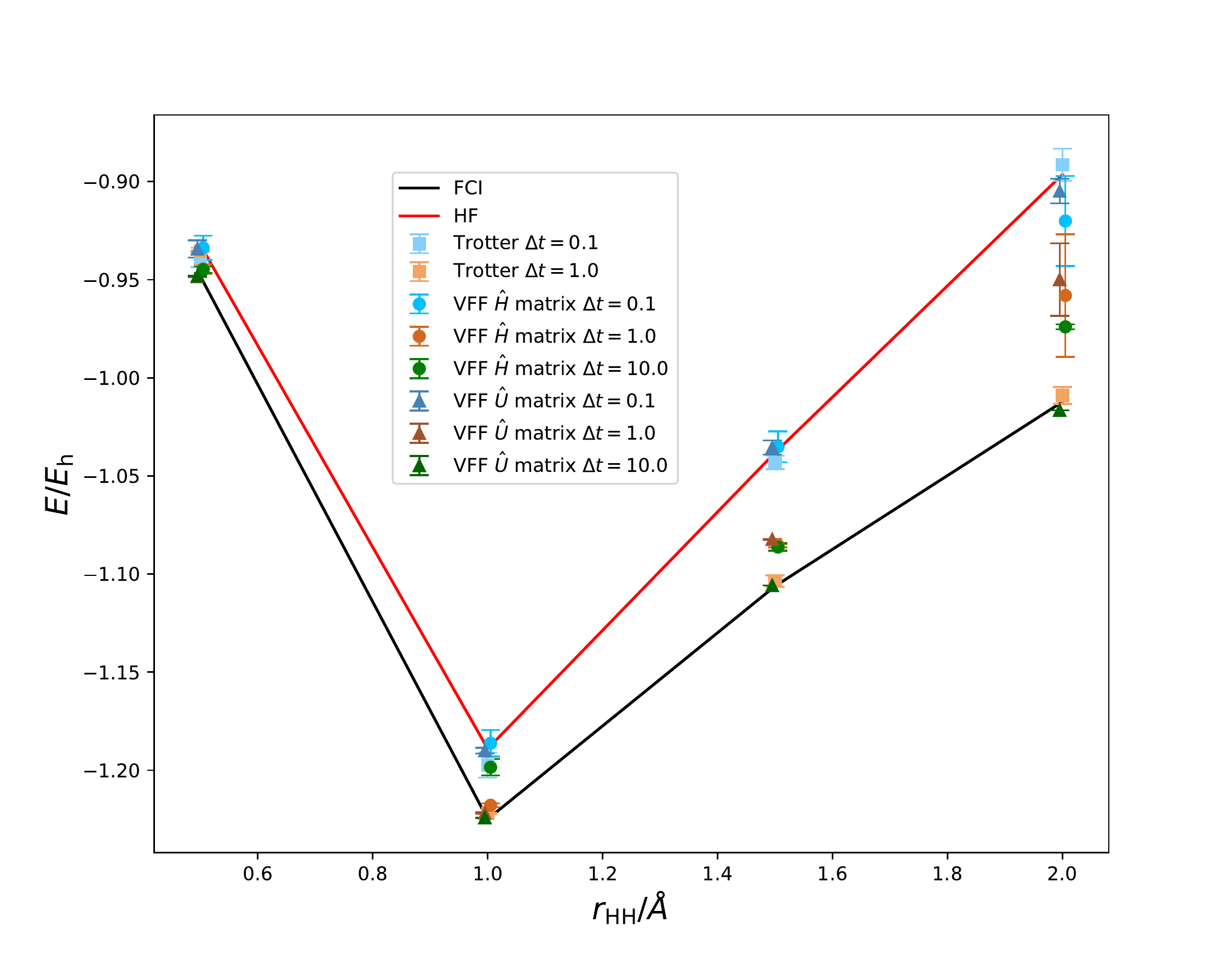}
    \includegraphics[width=0.5\textwidth, trim = 0 0 0 1cm, clip]{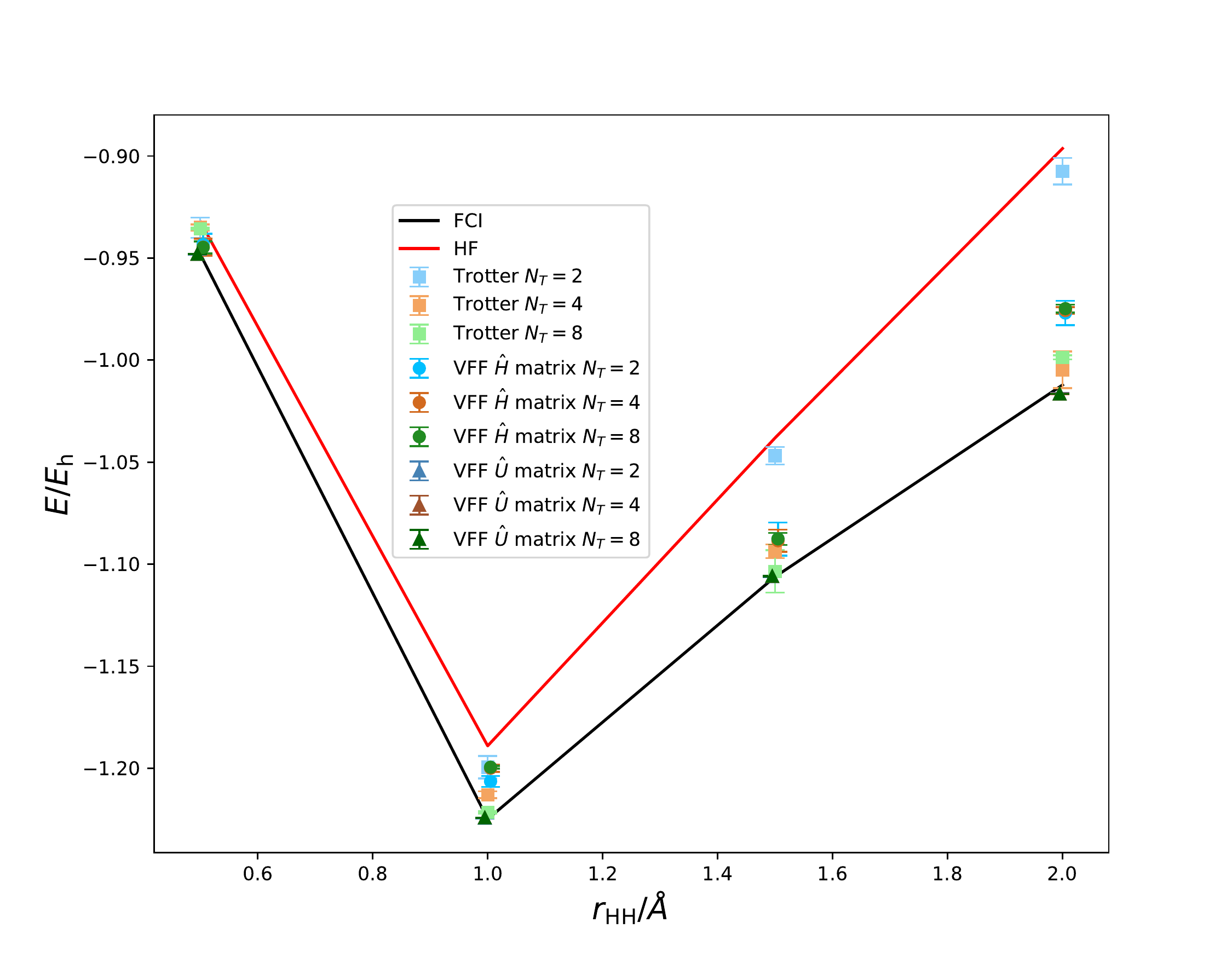}
    \caption{ \small Linear H$_3^+$ energy along the symmetric stretching mode in the STO-3G basis obtained from shot-based VQPE simulations, with or without the VFF approximation, using $N_T = 10$ and varying sizes of time grid (top) and $\Delta t = 1.0$ (VQPE) or $\Delta t = 10.0$ (VFF-VQPE) and varying numbers of basis states (bottom). Error bars were obtained by averaging results over multiple different simulations. VFF $\hat H$ matrix results are obtained using \cref{eq:gen_eig} and $\hat U$ matrix results with \cref{eq:gen_eig2}. For this system, VFF-based VQPE requires larger time-steps to converge than the underlying Trotter time-evolution. { In this case, the VFF optimisation leads to a very low cost function, suggesting VFF is a good approximation to the true time evolution, so} results obtained by diagonalising the $\mathbf{U}$ matrix may be used (triangles) and in general show better convergence than the comparable results obtained by Hamiltonian diagonalisation (circles). }
    \label{fig:H3}
\end{figure}

\begin{figure}[h!]
    \centering
    \includegraphics[width=0.5\textwidth, trim = 0 0 0 1.4cm, clip]{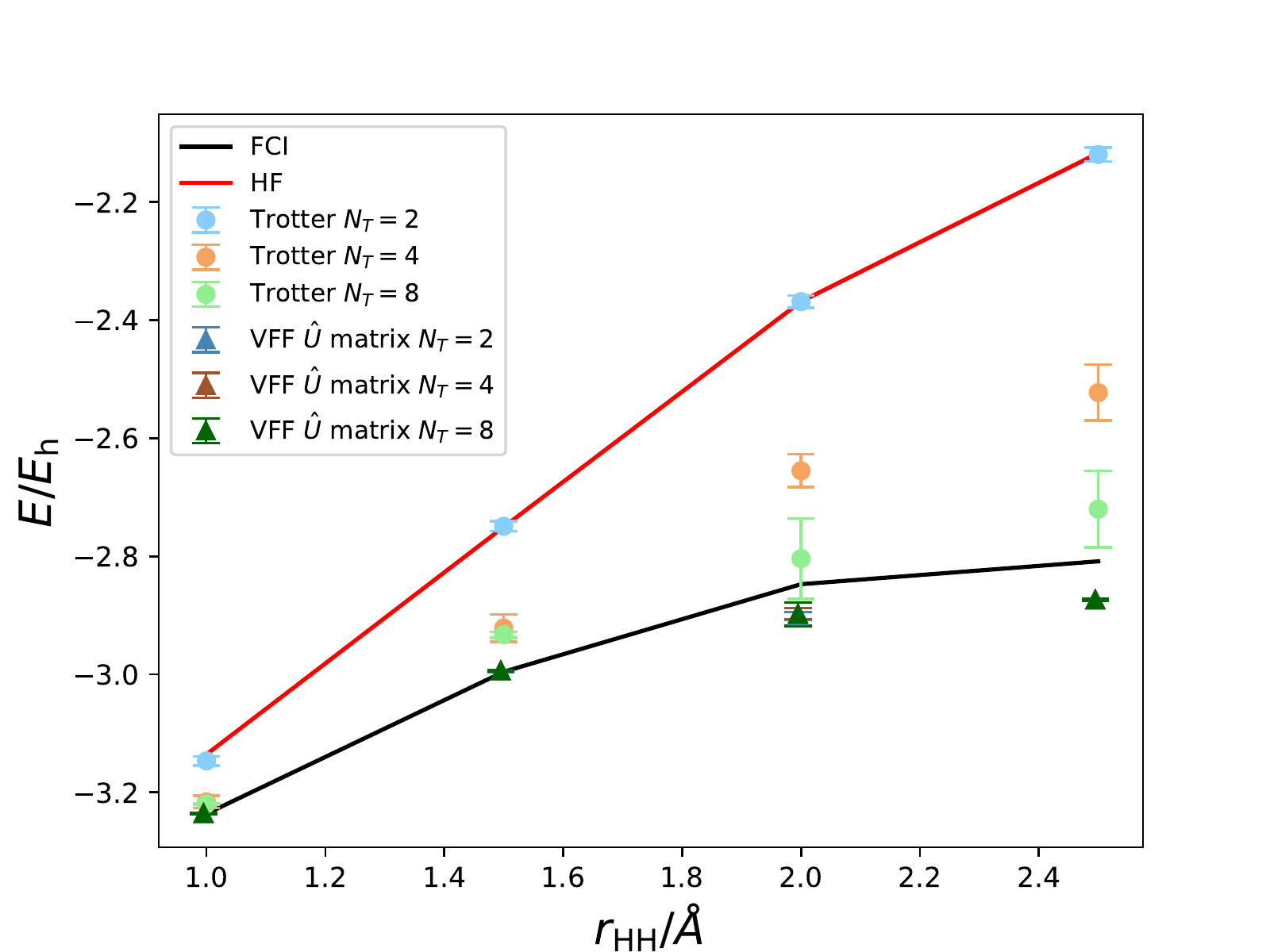}
    \caption{ H$_6$ chain energy along the symmetric stretch in the STO-3G basis obtained using $\Delta t=0.5$ (VQPE) and $\Delta t = 5.0$ (VFF-VQPE), with varying numbers of basis states.
    %For all VFF results and at $r_\mathrm{HH} = 0.5$ for VQPE, the energies shown are shifted by $2\pi$ from those directly obtained from the complex logarithm. 
    Error bars were obtained by averaging results over multiple different simulations. 
    %VFF $\hat H$ matrix results are obtained using \cref{eq:gen_eig} and 
    VFF $\hat U$ matrix results were obtained with \cref{eq:gen_eig2}. In this case, the diagonalisation of the unitary VFF matrix very efficiently recovers the correlation energy in a small number of steps.}
    \label{fig:H6}
\end{figure}

First, we note a few salient features of exact VPE in hydrogen systems. For H$_2$ and H$_3^+$ we considered a range of time-steps $\Delta t \in [0.05,2.0]$ and up to 10 time-steps in each case, with a representative subset of these results given in \cref{fig:H2,fig:H3}. For H$_6$ we found that the calculations were significantly more sensitive to the size of the time-step, with many values for $\Delta t$ giving unphysical energies. \cref{fig:H6} shows results obtained using the best attempted time-steps.

In all cases, as has been previously shown in the work of Klymko \textit{et al.}\cite{Klymko2022}, convergence with increasing number of time-evolved states is not smooth, but occurs in jumps. For example, in the bottom panels of \cref{fig:H2,fig:H3}, we observe that the $N_T=2$ values oscillate around the HF energy while for $N_T \geq 4$ the energy is converged to the FCI value. H$_6$ shows a similar, multi-stage process, although in this case the energy is not converged by $N_T = 10$ for stretched geometries (see \cref{fig:H6} { for results up to $N_T = 8$;  no significant changes were observed when increasing $N_T$ to 10}). Indeed, it is possible for the calculations not to converge at all and merely oscillate around the HF energy if the time-step is too small, as is the case for $\Delta t = 0.1$ in the H$_2$ and H$_3^+$ systems. 

All of these behaviours can be linked to the variation of the number of linearly independent states in the time-evolved basis. For H$_2$, the eigenbasis is two-dimensional, so a jump from the HF energy to the FCI value (see \cref{fig:H2}) is observed when time evolution successfully generates a second linearly independent state. For small $\Delta t$, ten steps are not enough to surpass the linear dependency threshold, so no change in energy is observed. If the linear independence threshold were lower, the second state may appear after fewer steps, but care needs to be taken to ensure no spurious states are generated by noise. In the noiseless, state-vector simulation of the time-evolution, one time step always generates a linearly independent state and therefore recovers the FCI energy regardless of the size of $\Delta t$.

\begin{figure}
    \centering
    \includegraphics[width=0.5\textwidth]{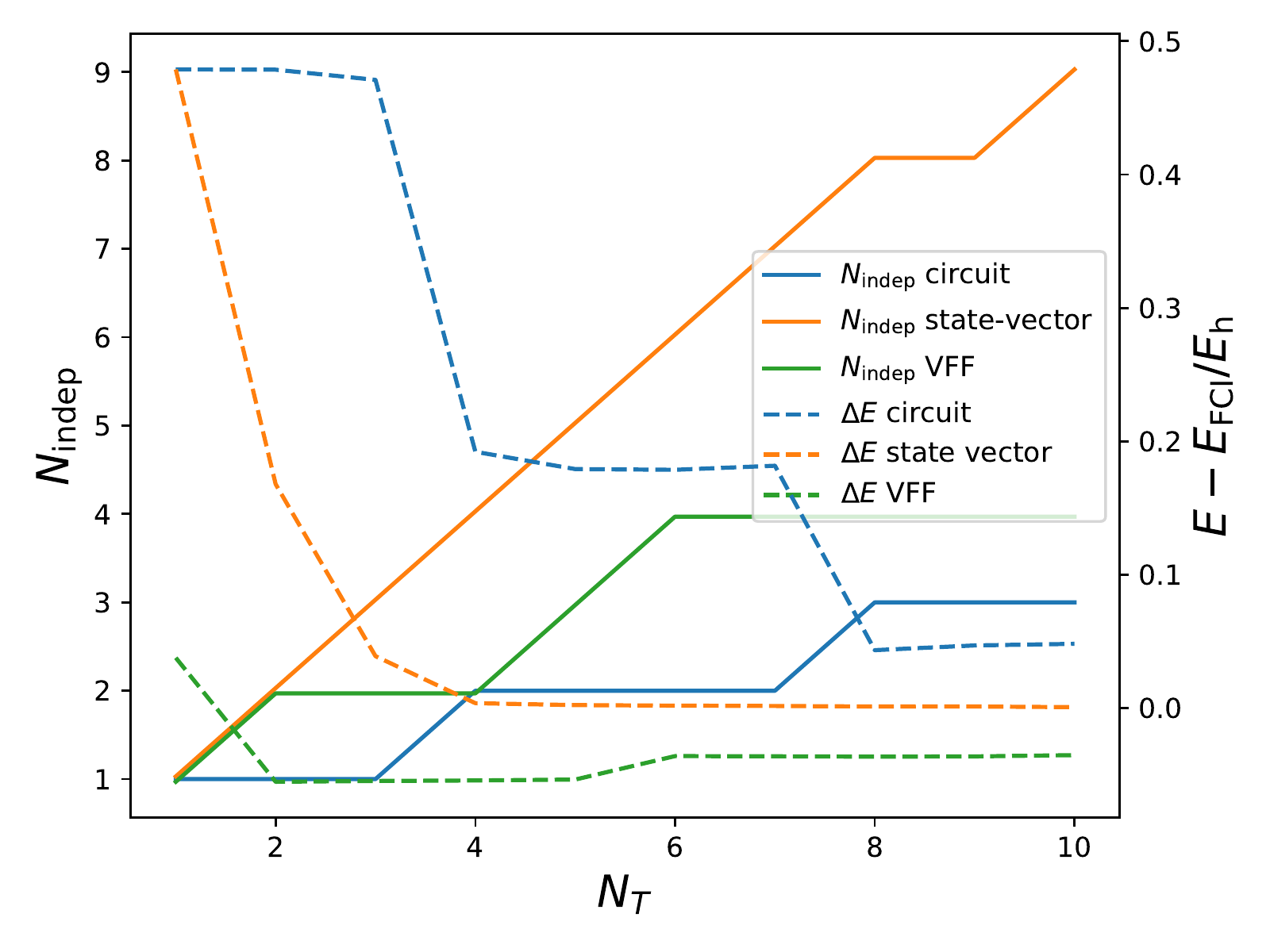}
    \caption{Dependence of the number of linearly independent states (solid lines) and error in the total energy (dashed lines) on the number of time-evolved basis states for the state-vector simulation (orange), a circuit-based Trotterised time-evolution (blue) and a circuit-based VFF evolution (green) of H$_6$ at $r_\mathrm{HH} = 2.0$ \r{A}, using $\Delta t = 0.5$. VFF generates new linearly independent states faster than the Trotter algorithm.}
    \label{fig:H6-lindep}
\end{figure}

Similar behaviour is observed for the H$_6$ chain in \cref{fig:H6-lindep}. At $\Delta t = 0.5$ and $r_\mathrm{HH} = 2.0$ \r{A}, the simulated state-vector evolution largely introduces one new linearly independent basis state for each time-evolved state, leading to fast, exponential convergence to the full CI solution. The circuit-based evolution introduces independent states much more slowly, with energies agreeing with the state-vector values using the same size of linearly independent basis.

\begin{table}
\footnotesize
\centering
\begin{tabular}{c|c|c|c|c}
&\multicolumn{2}{c|}{Trotterised evolution}&\multicolumn{2}{c}{VFF}\\
\cline{2-5}
System & Gates & CNOTs & Gates & CNOTs\\
\hline
H$_2$&112&34&57&24\\
\hline
H$_3^+$&671&300&85&36\\
\hline
H$_6$&15713&9638&1261&540\\
\hline
\end{tabular}
\caption{Number of total gates and CNOTs \textit{per trotterised time-evolution step} and in the full VFF time-evolution operator for the hydrogen systems in the STO-3g basis set.}
\label{tab:ngates}
\end{table}

 Therefore, when using the exact evolution four time\nobreakdash-steps are enough to recover the FCI energy, but significantly more are needed in the circuit-based evolution to get over the increased dependency threshold. This represents a substantial barrier to the application of the VQPE algorithm in this form on NISQ devices. 

 As can be seen from \cref{tab:ngates}, the number of gates required to evolve the system one { trotterised} step is large and grows rapidly with system size, quickly exceeding the depth of circuit that can be computed within the coherence times of current devices.\cite{Linke2017} While in this parametrisation, even one step would be challenging to compute, the number of steps could in principle be reduced by increasing the size of $\Delta t$ to more quickly surpass the dependency threshold. However, this would have the undesirable side-effect of increasing the error associated with the Trotter decomposition of the time-propagation operator. {Higher order Trotterisation could help reduce these errors,\cite{Childs2019} but would once again require deeper circuits.} Instead, we propose using a VFF-based approach to generate time evolved states.

\begin{figure}[h!]
    \centering
    \includegraphics[width=0.5\textwidth, trim = 0 0 0 1.4cm, clip]{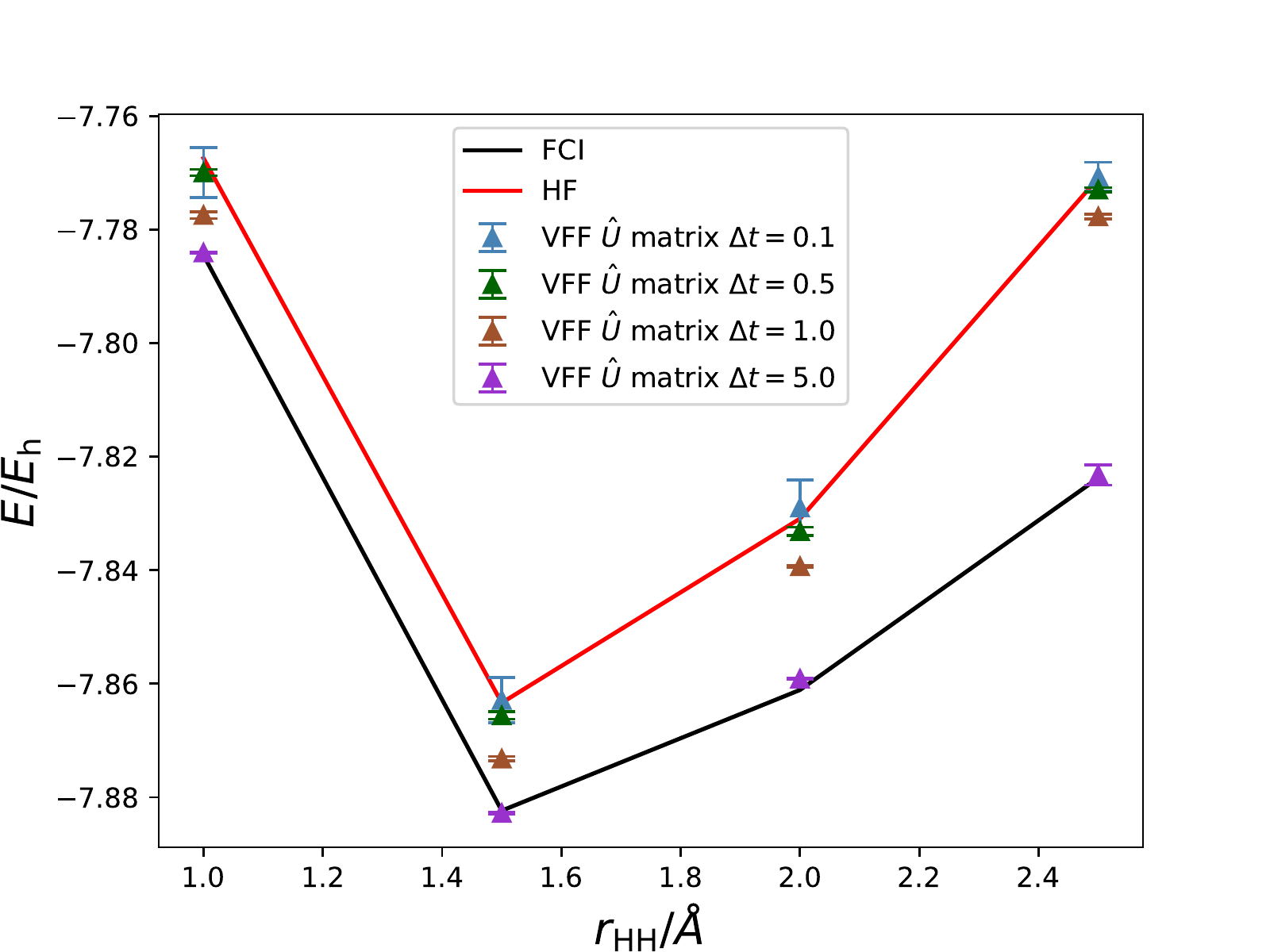}
    \includegraphics[width=0.5\textwidth, trim = 0 0 0 1.4cm, clip]{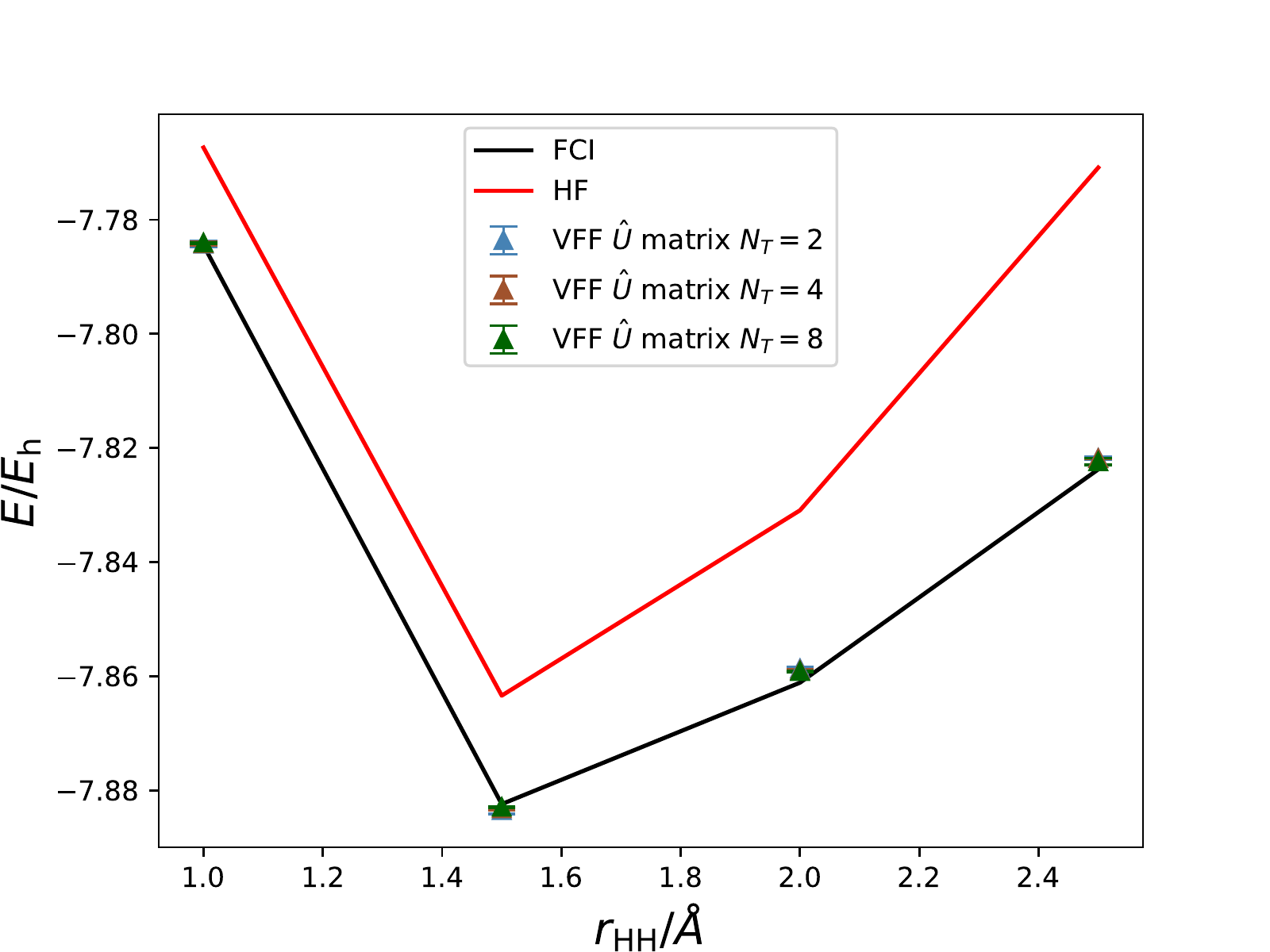}
    \caption{LiH energy along the dissociation curve in the STO-3G basis obtained by VFF-VQPE using $N_T = 10$ steps and varying $\Delta t$ (top) and $\Delta t = 5.0$ and varying $N_T$ (bottom). Error bars were obtained by averaging results over multiple different simulations. VFF $\hat H$ matrix results are obtained using \cref{eq:gen_eig} and $\hat U$ matrix results with \cref{eq:gen_eig2}. For large enough time-steps, convergence is very rapid with $N_T$.}
    \label{fig:LiH}
\end{figure}

By comparison, VFF introduces a constant-depth approximation to the time-evolution operator, with gate complexity comparable to a single Trotter step in the same system (see \cref{tab:ngates}, making the generation of VQPE states significantly more tractable. \Cref{fig:H2,fig:H3,fig:H6} present results obtained by diagonalising either the Hamiltonian or the $\mathbf U$ matrix { (constructed using \cref{eq:U_S_relation} rather than \cref{eq:U_def})} in the basis of VFF states. In general, we observe very similar behaviour to the pure VQPE algorithm, although for the more complex systems, larger time-steps are required in VFF than necessary in trotterised time-evolution. While for the smaller systems the convergence of VFF-VQPE appears to be similar or slower than that of exact VQPE, in H$_6$ the VFF-VQPE converges very quickly, provided a sufficiently large time-step is used.{ We also note that in this case, for large bond lengths, the energy is non-variational, due to the approximations employed in the computation of $\mathbf{U}$.}

Finally, we test the VFF approximation for VQPE in the LiH molecule in the STO-3G basis set, which has 12 qubits and 4 electrons. We only employ the unitary diagonalisation, as the linear cost of computing the overlap matrix leads to a significant acceleration in the algorithm for this size of system. The corresponding trotterised time-evolution approach becomes intractably slow for systems of this size. Unsurprisingly, we observe the same trends as in the H$_6$ chain --- a threshold in time-step size must be overcome for the method to converge, but once this has been achieved, convergence with the number of steps is very quick, with only one evolved state required for all geometries. This is extremely promising for the applicability of this method to larger molecules, although the required number of qubits is currently beyond the capacity of the quantum simulators employed in this work.

\subsection{Convergence properties of VFF-VQPE}

\begin{figure*}
    \centering
    \includegraphics[width=\textwidth, trim = 3cm 0 4.5cm 0, clip]{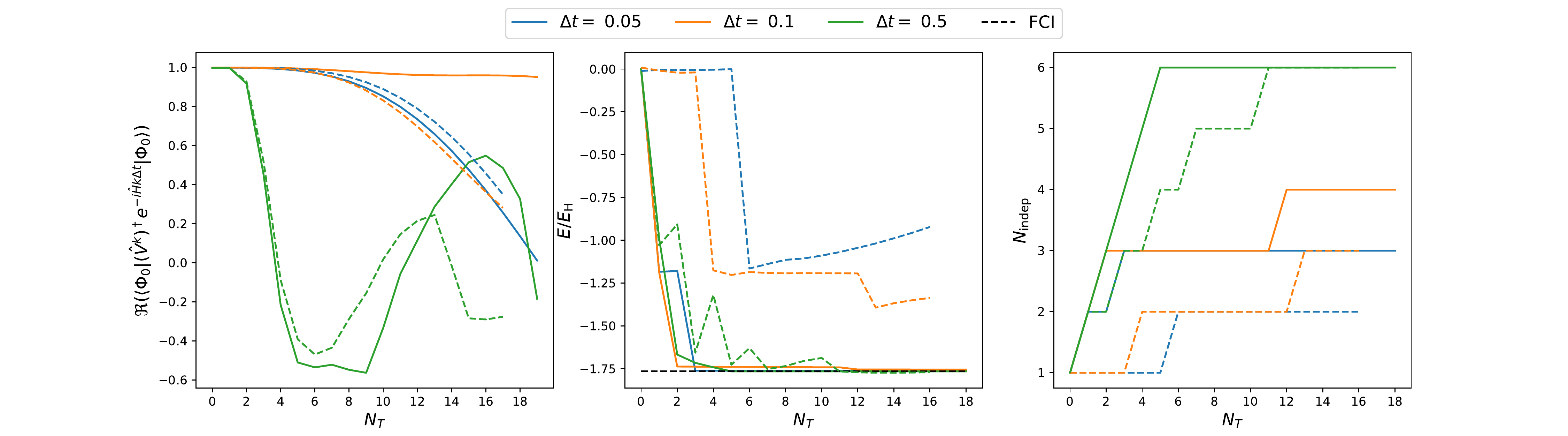}
    \caption{\small Results obtained by diagonalising the VFF-based VQPE Hamiltonian matrix for the Hubbard dimer with $U/t=0.5$. The VFF circuit was fitted to two time-steps, for $\Delta t \in \{0.05, 0.1, 0.5\}$ and a state-vector quantum circuit simulator (solid line) or shot-based simulator (dashed lines) was used. The left panel shows the overlap between the $k$-th VFF state and the corresponding time-evolved state. The value of the VQPE energy is given in the central panel and the number of linearly independent states is shown in the right panel. Fitting to at least two states is crucial to avoid triplet contamination and allow fast convergence to the singlet ground state. { Convergence for shot-based simulations is similar to state-vector simulations, but the presence of stochastic noise generates noticeable kinks in the overlap and energy curves, as well as slower introduction of new linearly independent states.}}
    \label{fig:state_VFF2}
\end{figure*}

We investigate the convergence properties of VFF-VQPE in more detail. { In order to do this, we focus on the results of VFF optimisation on two very simple systems - H$_2$ and the 2-site Hubbard model. We choose the latter because it is known to be accurately fast-forwardable\cite{Cirstoiu2020}, so we can attribute any errors clearly to the simplified optimisation methodology we employ here.} \cref{fig:state_VFF2} shows a series of state-vector VFF-based VQPE runs for the 2-site Hubbard model with $U/t = 0.5$. This system has a Hilbert space of 6 states, split into three symmetry sectors: $^1\Sigma_g^+$ (2 states), $^1\Sigma_u^+$ (1 state) and $^3\Sigma_u^+$ (3 states). These three sectors are not coupled by the Hamiltonian and therefore by the time-evolution operator, but they may be coupled by the VFF unitary matrix, since we have imposed no spatial symmetry constraints on $\hat W$. As long as VFF provides a good approximation to the true time-evolution operator, this coupling should be weak.

\begin{table}[h]
    \centering
    \begin{tabular}{c|c}
         $\min\Big(\Re{\braket{\psi|(V^k)^\dagger e^{-i\hat Hk\Delta t}|\psi}}\Big)$ & $|\Delta E_\mathrm{VQPE}|$ \\
         \hline
          0.844 & 0.1639\\
          \hline
          0.852 & 0.1128\\
          \hline
          0.95570 & 0.0409\\ % negative
          \hline
          0.95744 & 0.0426\\ %negative
          \hline
          0.95748 & 0.0131\\
          \hline
    \end{tabular}
    \caption{ Absolute error in the VQPE energy of some example calculations for the Hubbard dimer obtained by diagonalising the $\mathbf{U}$ matrix as a function of minimum overlap between the VFF and time-evolved states. We find the error to generally decrease as agreement between the two increases.}
    \label{tab:VFF_unitary}
\end{table}

In all cases, $\hat D$ was truncated to $m=1$ and $\hat W$ was parametrised by a single-layer symmetry-preserving ansatz, leading to a circuit with a depth of 57 gates, of which 24 are CNOTs. This is a significant improvement from the original Trotter time-evolution, particularly as the depth of this circuit remains constant for any number of steps. More complex parametrisations were attempted, but no significant change to the results was observed.{ The error relative to true time evolution increases with number of steps, but this does not correlate with larger errors in the VPE results.} In fact, as expected from the previous discussion of direct Hamiltonian diagonalisation in the Krylov subspace, the quality of VFF agreement with the true time-evolved states has little correlation with quality of convergence in general, with the number of linearly independent states generated by repeated application of the unitary being the dominating factor in the quality of the obtained energy by far.

In state-vector simulation, all calculations converge to the ground state, even when spanning fewer independent states than the 6 present in FCI. The behaviour of shot-based simulation in \cref{fig:state_VFF2} is qualitatively similar to that of the state-vector simulations, although the effects of random noise are now clearly visible in behaviour such as the unphysical upwards drift in the $\Delta t = 0.05$ energy. Nevertheless, larger time-steps lead to successful convergence to the ground state.

{ As discussed in the introduction, for sufficiently good agreement between the true time-evolved states and their VFF counterparts, it should be possible to obtain estimates of the energy eigenvalues from an approximate $\mathbf{U}$ matrix.} We consider the $\Delta t = 0.1$ case for the Hubbard model in \cref{fig:state_VFF2} and attempt to construct the $\mathbf{U}$ matrix from the overlap. \cref{tab:VFF_unitary} gives the results of the diagonalisations of a series of such matrices, { each obtained from a different optimisation of the VFF parameters for the same underlying system}.
We find that, on a large scale, better overlap between the VFF and true time-evolved states leads to better energy values, while close to perfect agreement between the states, the energy oscillates about the true FCI value. As can be seen from \cref{fig:H2,fig:H3,fig:H6}, some oscillation is also present in the shot-based results based on the true time-evolved states. These results are promising and suggest that, given a reliably good Hamiltonian approximation, the VFF-based VQPE method can also be applied with only linear cost in the number of basis functions used. Further exploration into how to guarantee the quality of the approximation is warranted.

We investigate the possibility of diagonalising the unitary matrix for hydrogen systems as well. While general physical Hamiltonians cannot necessarily be fast forwarded,\cite{Cirstoiu2020,Atia2017} { that is to say their time-evolution cannot be simulated with a constant-depth circuit,
we find that, for the molecules considered here it is possible to obtain a good VFF approximation to the time-evolution. For example, \cref{fig:H2_VFF} shows the results of VFF optimisation for H$_2$.} In most cases the error in the overlap with the true trotterised state is less than $10^{-6}$, so this system can be well treated with the unitary approach. As seen from the quality of unitary results in \cref{fig:H2,fig:H3,fig:H6}, these trends continue beyond the simple H$_2$ case.

Somewhat surprisingly, results obtained from the diagonalisation of the $\mathbf{U}$ matrix have smaller error bars than those obtained from the Hamiltonian. This may be due to the fact that the $\mathbf{U}$ and $\mathbf{S}$ matrices are obtained from the same set of measurements, leading to correlated errors in the two which may cancel in the overall result.

%% file: sections/conclusion.tex
\section{Conclusions} \label{sec:conclusion}

We have presented a circuit-based implementation for the VQPE algorithm based on
Trotterised time propagation and the Hadamard test. This algorithm was implemented
successfully on both state-vector and shot-based quantum simulators and we find that the method is
successful in finding the ground state of the H$_2$ and H$_3^+$ systems, but
requires a prohibitively large number of one- and two-qubit gates to converge
for H$_6$ or LiH. This is an inevitable consequence of Trotterised time-evolution and
would prevent this algorithm from finding significant use of NISQ devices.

\begin{figure*}
    \centering
    \includegraphics[width=\textwidth, trim = 3cm 0 4.5cm 0, clip]{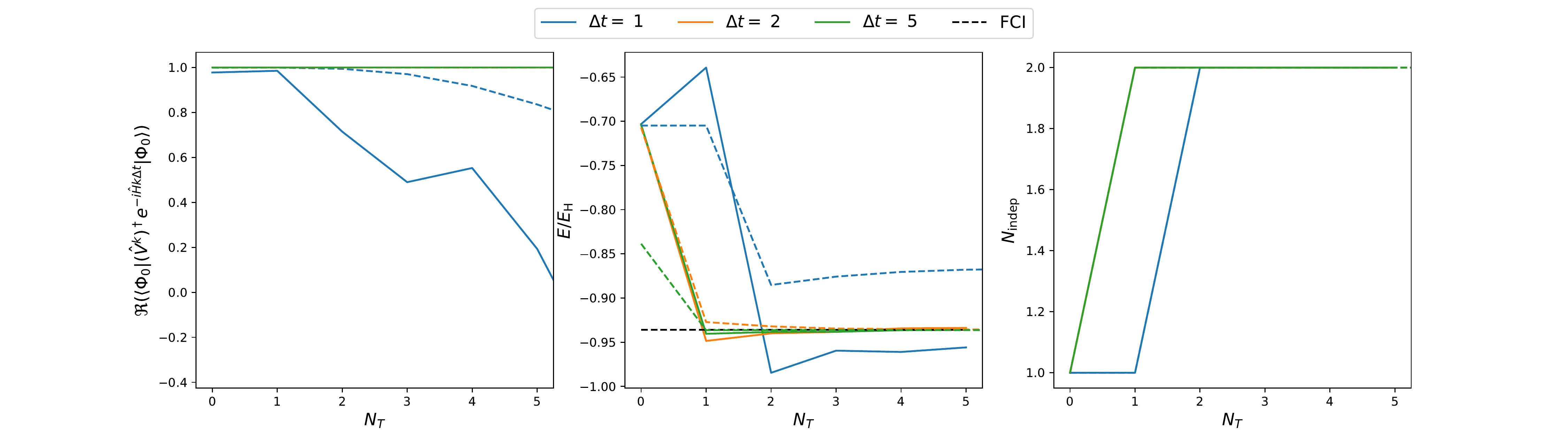}
    \caption{\small Results obtained by diagonalisation of the VFF-based VQPE Hamiltonian matrix (solid lines) or $\mathbf{U}$ matrix (dashed lines) for H$_2$ at $r_\mathrm{HH} = 2.5$\AA. The VFF circuit was fitted to two time-steps, for $\Delta t \in \{1, 2, 5\}$ and a shot-based quantum circuit simulator was used. {The orange line corresponding to  $\Delta t = 2$ perfectly overlaps with the green $\Delta t = 5$ line in the left and right panels.} The left panel shows the overlap between the $k$-th VFF state and the corresponding time-evolved state. The value of the VQPE energy is given in the central panel and the number of linearly independent states is shown in the right panel. In this case, we observe a correlation between the quality of the overlap between the fast-forwarded state and the true time-evolved state, and the rate of convergence of VQPE.}
    \label{fig:H2_VFF}
\end{figure*}

We propose an approximation to the VQPE algorithm using the VFF approach, which
reduces the circuit depth of the time-evolution circuits to roughly that of a
single Trotter step. The approximation provides a sensible Krylov subspace basis
for Hamiltonian diagonalisation even when its fidelity to the true time-evolved
basis is low. Our particular choice of VFF decomposition natively preserves
electron number but not $m_s$ values or spatial symmetry, unlike true time-evolution, so potentially
requires artificially many linearly independent states to reach the true ground state. This
contamination can be reduced by using more time-evolved states in the cost function
for VFF optimisation, but it is highly desirable to develop approximations that natively implement
more of the symmetries of the time-evolution operator, thereby guaranteeing the preservation of the support
space and therefore imposing a maximum on the number of linearly independent states needed for convergence.

In general, VQPE based on VFF states requires construction of the Hamiltonian matrix,
so a quadratic number of calls to the quantum processor. However, where 
$\braket{\psi|(V^k)^\dagger e^{-iHk\Delta t}|\psi}$ is close to unity across the entire
range of VFF states considered, diagonalising the matrix $\mathbf{U}$ defined in \cref{eq:U_S_relation}
gives a good approximation of the true Hamiltonian eigenstates. In this case, VFF can be used
to reduce required quantum circuit depth while maintaining the linear cost of matrix generation
of exact VQPE. We find that this is highly successful even for more complex molecular systems like H$_6$ and LiH.

The number of gates required for trotterised time-evolution (see \cref{tab:ngates}) makes the algorithm intractable on near-term hardware. However, the significant depth reduction in the VFF approach makes it significantly more amenable to a near-term implementation, with recent experiments on the Quantinuum H1 quantum computer successfully employing circuits with up to 900 CNOTs.\cite{yamamoto2023demonstrating} The large number of measurements required even by the linear-scaling version of VQPE made it impossible to run experiments using this slow trapped-ion architecture, but the algorithm may be more feasible to execute on hardware with faster gate times, such as machines based on atomic Rydberg states.\cite{Jaksch2000}  

%% file: sections/acknowledgements.tex
%\section*{Acknowledgements} \label{sec:acknowledgements}

%% file: sections/appendices.tex
\appendix
\section{Quantum Phase Estimation} \label{sec:qpe}
Given a unitary operator $\hat U$ and one of its eigenfunctions $\ket{\Psi}$ such that $\hat U\ket{\Psi} = e^{i2\pi\theta}\ket{\Psi}$, QPE can be used to estimate the value of $\theta$.

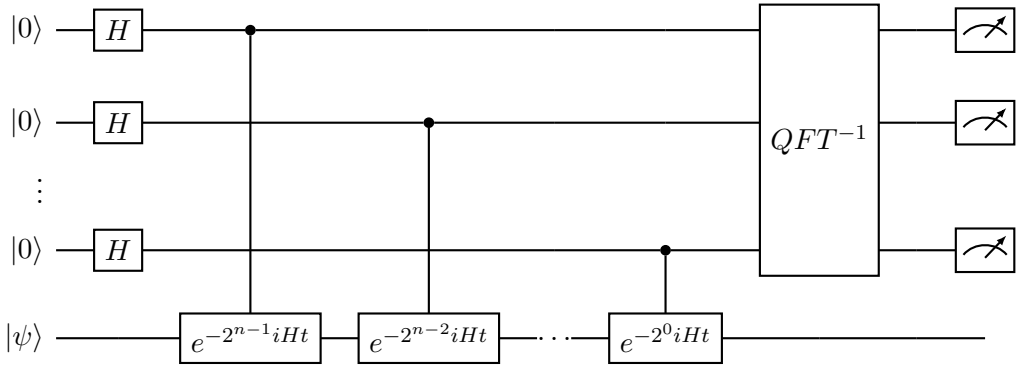
\begin{figure*}
    \centering
\begin{quantikz}
\lstick{$\ket{0}$} &\gate{H}  &\ctrl{4} &\qw   &\qw &\qw& \gate[4, nwires=3]{QFT^{-1}}&\qw &\meter{}\\
\lstick{$\ket{0}$} &\gate{H}   &\qw&\ctrl{3}   &\qw &\qw  &&\qw &\meter{}\\
\lstick{\vdots} &&&&&&&&&& \\
\lstick{$\ket{0}$} &\gate{H}  &\qw &\qw &\qw  & \ctrl{1} &&\qw&\meter{}\\
\lstick{$\ket{\psi}$} &\qw & \gate{e^{-2^{n-1}iHt}} &\gate{e^{-2^{n-2}iHt}}  &\qw\ldots &\gate{e^{-2^0iHt}}&\qw&\qw &\qw\\
\end{quantikz}
    \caption{QPE circuit.}
    \label{fig:qpe_circuit}
\end{figure*}

For a hermitian Hamiltonian $\hat H$, the corresponding time-evolution operator $\hat U = e^{-i\hat Ht}$ is unitary and may be used in QPE, using the circuit given in \cref{fig:qpe_circuit}. Before the inverse quantum Fourier transform (QFT), the generated wavefunction is given by
\begin{equation}
\ket{\Psi_\mathrm{QPE}} = \frac{1}{2^{n/2}}\sum_{j = 0}^{2^n - 1} e^{-i\hat Htj} \ket{j}\ket{\psi}.
\label{eq:qpe_wfn}
\end{equation}
where $\ket{j}$ is the qubit product state corresponding to the binary encoding of $j$. Taking the  inverse quantum Fourier transform gives
\begin{widetext}
    \begin{equation}
\mathrm{QFT}^{-1}[\ket{\Psi_\mathrm{QPE}}] = \frac{1}{2^{n/2}}\sum_{k = 0}^{2^n - 1} \ket{k}\Big(\sum_{j = 0}^{2^n - 1}e^{-itj(\omega_k  + \hat H)} \ket{\psi}\Big),
\label{eq:qft}
\end{equation}
\begin{equation}
\mathrm{QFT}^{-1}[\ket{\Psi_\mathrm{QPE}}] = \frac{1}{2^{n/2}}\sum_{k = 0}^{2^n - 1} \ket{k}\Big(\sum_{j = 0}^{2^n - 1} e^{-itj(\omega_k  + E)}\ket{\psi}\Big),
\label{eq:qft2}
\end{equation}

\end{widetext}

where $\omega_k = 2\pi k /(2^n t)$ and \cref{eq:qft2} can be obtained from \cref{eq:qft} if $\ket{\psi}$ is an eigenfunction of $\hat H$, 
\begin{equation}
    \hat H \ket{\psi} = E \ket{\psi}.
\end{equation}

$\mathrm{QFT}^{-1}[\ket{\Psi_\mathrm{QPE}}]$ peaks around $\omega_k = - E$, so measurement is highly likely to give an $n$-bit integer approximation of $k~=~- 2^n t E/2\pi$.

Provided a guess wavefunction $\ket{\tilde{\psi}}$, QPE will successfully generate the eigenvalue $E$ with probability proportional to $\abs{\braket{\tilde{\psi}|\psi}}^2$.\cite{Aspuru-Guzik2005} As such, if one has access to a wavefunction with good overlap with the ground state of a quantum system, QPE can be used to estimate the true ground state energy with high probability. For quantum chemical systems, the Hartree--Fock (HF) wavefunction may often be good enough, but techniques such as adiabatic state preparation\cite{Farhi2000,Farhi2001} may be used to improve upon it.

\subsection{Comparison to Variational Quantum Phase Estimation}

Consider $N_T = 2^n - 1$ time-evolved states using an evenly-spaced time-grid, where $n$ is the number of measurement qubits in the QPE circuit. The QPE wavefunction in \cref{eq:qpe_wfn} can be expressed in this basis as
\begin{equation}
\ket{\Psi_\mathrm{QPE}} = \frac{1}{\sqrt{N_T + 1}}\sum_{j = 0}^{N_T}\ket{j}\ket{\phi_{j,0}}.
\end{equation}

Similarly, the result of the inverse QFT is given by
\begin{widetext}
   \begin{equation}
    \mathrm{QFT}^{-1}[\ket{\Psi_\mathrm{QPE}}] = \frac{1}{N_T + 1} \sum_{k=0}^{N_T} \ket{k} \Big(\sum_{j=0}^{N_T}e^{i\omega_k t_j} \ket{\phi_{j,0}}\Big)= \frac{1}{\sqrt{N_T + 1}}  \sum_{k = 0} ^{N_T} \ket{k}\ket{\omega_k}.
\label{eq:qpe_vpe}
\end{equation} 
\end{widetext}
where 
\begin{equation}
    \ket{\omega_k} = \frac{1}{\sqrt{N_T + 1}}\sum_{j = 0}^{N_T} e^{i \omega_k t_j} \ket{\phi_{j,0}}
\end{equation}
form a Fourier basis.

We can alternatively consider these as particular linear combinations of the VQPE time-evolved states, $\ket{\phi_{j,0}}$. 
In order to obtain these same states from the VQPE algorithm, the Fourier basis would need to correspond to the solutions to the corresponding VQPE generalised eigenvalue problem. In that case, the resulting VQPE energy would also correspond to the QPE estimate, but this diagonality condition is not guaranteed to hold in general. Considering instead a fixed-error estimate of the eigenspectrum, Klymko \textit{et al} \cite{Klymko2022} estimate that the ratio of number of exponential operator applications needed by each algorithm scales asymptotically as
\begin{widetext}
\begin{align}
        \frac{N_\mathrm{VQPE}}{N_\mathrm{QPE}} &\in \widetilde{O}\Big[\frac{(N_T\sum_j|h_j|)^{9/2}||S^{-1}||^5t_\mathrm{max}^{3/2}\min_N |\phi_N^0|^2}{\epsilon} \Big]
\end{align}
\end{widetext}
where $h_j$ are the coefficients of different Pauli terms in the Hamiltonian, $t_\mathrm{max}$ is the maximum step size used in VQPE and $\epsilon$ is the desired error in the eigenvalues. Which algorithm is more efficient therefore depends significantly on the underlying Hamiltonian and initial state.